\documentclass[aps,pra,twocolumn,showpacs,groupedaddress]{revtex4-1}

\usepackage{stmaryrd}
\usepackage{amssymb}
\usepackage{amsmath}
\usepackage{amsfonts,dsfont,hyperref}
\usepackage{color}
\usepackage{epsfig}
\usepackage{bm}
\usepackage{rotating}
\usepackage{mathrsfs}
\usepackage{setspace}

\newcommand{\be}{\begin{equation}}
\newcommand{\ee}{\end{equation}}
\newcommand{\bea}{\begin{eqnarray}}
\newcommand{\eea}{\end{eqnarray}}
\newcommand{\ket}{\rangle}
\newcommand{\bra}{\langle}
\newcommand{\T}[2]{\textsf{#1#2}}
\newcommand{\C}[1]{\mathcal{#1}}

\newcommand{\I}{\mathds{1}}

\begin{document}

\title{Choi states, symmetry-based quantum gate teleportation, and stored-program quantum computing}

\author{Dong-Sheng Wang}
\affiliation{Institute for Quantum Computing and Department of Physics and Astronomy, \\ University of Waterloo, Waterloo, ON, Canada}

\begin{abstract}
The stored-program architecture is canonical in classical computing,
while its power needs more investigations for the quantum case.
We study quantum information processing with stored quantum program states,
i.e., using qubits instead of bits to encode quantum operations.
We develop a stored-program model based on Choi states,
following from channel-state duality, and
a symmetry-based generalization of deterministic gate teleportation,
which can be employed to compose Choi states.
Our model enriches the family of universal models for quantum computing,
in particular, the measurement-based models,
and can also be employed for tasks including quantum simulation and communication.
\end{abstract}
\date{\today}
\maketitle


\section{Introduction}

Quantum computing has been a fascinating field~\cite{Fey82,Shor94,NC00} and
quantum computers are believed to be very powerful
and becoming more foreseeable~\cite{LJL+10,AAB+19}.
As quantum systems,
there is the inevitable issue of how to control or monitor a quantum computer without disturbing its computation.
The external controllers might be quantum and entangled with the computer too much,
or too classical to yield effective control.
This has been discussed decades ago in the model of quantum Turing machine~\cite{Mye97}.
In the so-called gate-array or circuit model,
which is the commonly employed model and
in which computation is done by a sequence of unitary gates followed by measurements,
each gate may be induced by the interaction of qubits with external control fields and
is classically monitored such as its timing and location.
Furthermore, the sequence of gates, which compose an algorithm or program,
is stored as data in classical computers,
and there is a classical algorithm that designs such a sequence of gates,
such as the Solovay-Kitaev algorithm~\cite{NC00,DN06}.

In this work, we study one such quantum-classical interface issue,
which is whether or how to store quantum circuits as quantum programs.
The analog in the classical setting is the von Neumann architecture,
also known as stored-program architecture.
A major advantage is that a computer can store various programs as data,
whose action can be executed or simulated at any later stage on demand.
In the era of quantum information science,
such a stored-program paradigm is highly desirable
and turns out to be nontrivial~\cite{Mye97,NC97,Shi02,KPP19,SBZ19}.
Nielsen and Chuang~\cite{NC97} proved that if the processor is
independent of the input quantum programs and is fixed,
then it is not possible to achieve universality.
In particular, the seemingly inevitable entanglement between quantum program
and processor causes difficulty,
and it might be too restrictive to keep the processor fixed.

It turns out there are quantum computing models with a certain stored-program features,
and they are known as teleportation-based or measurement-based quantum computing~\cite{GC99,RB01,BBD+09},
although motivated from different perspectives.
For instance, the computing on cluster states,
which are entangled states built up with Hadamard gates and controlled-not gates~\cite{RB01},
is by local projective measurements in
bases that are determined by previous measurement outcomes.
A sequence of gates in the circuit model is simulated by a sequence of measurements.
We may pull back the measurement bases as initial states,
leading to different program states for different computations.
In the scheme based on Bell measurement~\cite{GC99},
gates, hence program states, are sorted in the levels of Clifford hierarchy and can be teleported
with byproduct correction operators of one-level lower.
However, in these models the program states are of particular types,
and our study in this work serves as a modification or generalization of them.

Quantum data are commonly stored as quantum states of multiple qubits.
We employ the channel-state duality, also known as Choi–Jamiołkowski isomorphism~\cite{Jam72,Cho75,BZ06},
to convert quantum operations,
which are the quantum programs,
to states, which are called program states.
Such program states are also called Choi states in this work.
For instance, the Choi state of a unitary operator $U$ is a vector by shuffling
each column of $U$ to the first column one after another.
The channel-state duality is
 a fundamental relation in quantum theory yet
only has limited applications so far such as quantum tomography~\cite{NC00}.

We develop a stored-program model based on Choi states and also
a generalization of gate teleportation~\cite{BBC+93,GC99}.
The model contains a program and a processor,
with Choi states serving as the programs
and a symmetry-based gate teleportation (and also a readout scheme) as the processor.
The processor will in general depend on the input program state.
We consider general program states,
namely, states for completely positive and trace-preserving mappings,
i.e., quantum channels~\cite{BZ06,NC00}.
By reformulating gate teleportation in the language of matrix-product states~\cite{PVW+07},
we treat gates as symmetry operators on tensors
instead of operators on states~\cite{Wang19b}.
This leads to our symmetry-based gate teleportation,
which is able to reveal the computational powers of Choi states.

We mainly study three primary problems for our stored-program model:
the preparation of a program state,
the composition of program states,
and the execution of the effects of the encoded programs.
A mixed program state will be stored as a set of extreme program states and random bits
based on extreme channels~\cite{WS15,Wang16},
which is more efficient in terms of circuit costs than the standard dilation method.
Given many program states,
it might be desirable to compose them into a larger one for many purposes,
e.g., to compose or concatenate algorithms together,
and this turns out to be nontrivial and we develop the
symmetry-based gate teleportation to achieve the composition.
Finally, we also discuss two different schemes for readout,
i.e., the execution of channels from Choi states.

Our model can also be adopted for other tasks,
such as quantum channel simulation and dissipative quantum computing~\cite{VWC09,KBG+11}.
In the setting of quantum communication,
our scheme can also be viewed as an extended form of quantum repeaters
that are important to build quantum networks~\cite{BDC+98,Kim08}.
We present our work in the following sections
depending on their technical sophistication.
In section~\ref{sec:pre},
we review basics for quantum channels, teleportation,
and their connection with symmetry..
In section~\ref{sec:cr},
we discuss methods for channel-execution from Choi states.
Then in section~\ref{sec:comp} we develop the symmetry-based gate teleportation
and the scheme for composition of Choi states.
In section~\ref{sec:extr} we present a method for the preparation of
Choi program states via generalized-extreme channels.
Then we discuss some applications of the model we develop in section~\ref{sec:app},
in particular, the stored-program quantum computing.
We conclude with some perspectives in section~\ref{sec:conc}.

\section{Preliminary}
\label{sec:pre}

We start from a brief survey of quantum channels~\cite{BZ06,NC00}
and the symmetry point of view of teleportations~\cite{Wang19b}.
This leads to our generalization of gate teleportation and the computing scheme based on Choi states.

\begin{figure}
  \centering
  \includegraphics[width=.48\textwidth]{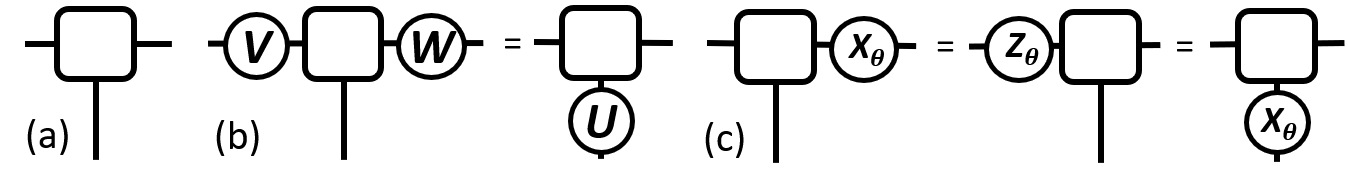}
  \caption{(a) A 3-leg tensor with a vertical physical leg $p$, the left leg $l$, and the right leg $r$.
  The input (output) is carried by the left (right) leg $l$ ($r$).
  The Kraus operators are represented by the box.
  (b) A symmetry of a 3-leg tensor.
  (c) The symmetry of the Z-bit teleportation.}\label{fig:tensor}
\end{figure}

We consider finite-dimensional Hilbert space $\C H$ and the set of density operators on it $\C D(\C H)$.
A quantum channel $\C E$ acting on $\C D(\C H)$ can be represented by a set of Kraus operators $\{K_i\}$
such that $\sum_i K_i^\dagger K_i=\I$, for dim$\C H:=d$.
It can be viewed as an isometry $V=\sum_i |i\ket K_i$, with $V^\dagger V=\I$ but $VV^\dagger \neq \I$.
It can also be viewed as a three-leg tensor: $i$ as the index for the physical space,
and each $K_i$ acts on the logical space~\footnote{The `logical' space is also known as `virtual' or `correlation' space.
The `physical' space here is also an ancillary space.}.
See Fig.~\ref{fig:tensor} (a) for a tensor.
The input (output) for the channel is usually carried by the left (right) leg $l$ ($r$),
and the vertical physical leg $p$ labels different Kraus operators.

We say a channel $\C E$ has symmetry group $G$ when
\be \sum_{j} U^g_{ij} K_j=W^g K_i V^g, \forall g\in G, \ee
for unitary representations $V$, $W$, and $U=(U_{ij})$.
See Fig.~\ref{fig:tensor} (b).
The symmetry is called `global' when $W=V^\dagger$ since in this case
it can be extended to composition of channels;
otherwise, the symmetry is called a `local' or `gauge' symmetry.
A gauge symmetry can induce a smaller global symmetry under composition.
With no surprise, a sequence of tensors will lead to a matrix-product state~\cite{PVW+07}.

It turns out teleportation can be understood from the point of view of symmetry~\cite{WSR17,Wang19b},
and then be generalized.
A simple example is the 1-bit teleportation~\cite{ZLC00},
which can be an X-bit or equivalently a Z-bit one with $X$ or $Z$ as `byproduct' operators that can be corrected.
Note we use $X,Y,Z$ to denote Pauli matrices.
For instance, the Z-bit teleportation is
\be M^s_{X_1} CX_{12} |\psi\ket_1 |0\ket_2 = Z_2^s |\psi\ket_2, \ee
for two qubits with initial states $|\psi\ket$ and $|0\ket$,
the controlled-not gate $CX$ with the 1st qubit as control,
and final measurement of Pauli $X$ on the 1st qubit with outcome $s=0,1$ for projections $|\pm\ket$,
and the arbitrary qubit state $|\psi\ket$ is teleported from the 1st qubit to the 2nd one.
We can swap the two qubits before the final measurement and instead measure the 2nd qubit,
then the process is a channel, with Kraus operators $H$ and $HZ$ with equal probabilities,
on the 1st qubit if we mix the two outcomes together,
for $H$ as Hadamard gate.
Hence the Z-bit teleportation is the tensor $\{HZ^i\}$, $i=0,1$.
The usual teleportation, which is 2-bit, is obtained as a sequence of two Z-bit teleportation
(or equivalently X-bit teleportation) with $\{X^iZ^j\}$ for $X^i=HZ^iH$.

Now gate teleportation follows from the symmetry of them.
The 1-bit teleportation has the $U(1)$ gauge symmetry:
a rotation $X_\theta:=e^{i\theta X}$ on the physical leg $p$
is equivalent to $X_\theta$ on the right logical leg $r$ or $Z_\theta$ on the left leg $l$.
See Fig.~\ref{fig:tensor}, (c).
The 1-bit teleportation also has a $Z_2$ symmetry: $Z_p X_l Z_r$,
which induces a $Z_2\times Z_2$ global symmetry for the 2-bit teleportation,
which has Pauli operators as the projective representation of it~\cite{Wang19b}.
This symmetry point of view also applies to qudit cases
with Hadamard replaced by Fourier transform operators.
This forms the foundation for the computing scheme with cluster states~\cite{RB01,WSR17}.

It is straightforward to see that
the symmetry can be significantly enlarged to $SU(d)$ for general qudit cases
when the identity operator is absent from the tensor of teleportation.
The symmetry relation reads
\be \sum_{j}T_{ij} P_j= V^\dagger P_i V \ee
for $V\in SU(d)$, $T_{ij}=\T tr(P_i V P_j^\dagger V^\dagger)/d$,
and $P_i$ as generalized Pauli operators, namely,
the hermitian Gell-Mann operators or the unitary Heisenberg-Weyl operators,
which both work
and we find there is no need to restrict to either one of them.
The difference between them is a change of basis.
The matrix $T:=(T_{ij})$ is actually the affine representation~\cite{BZ06} of the gate $V$,
and it is certainly unitary.
We will use this large symmetry to generalize gate teleportation below.

\begin{figure}
  \centering
  \includegraphics[width=.45\textwidth]{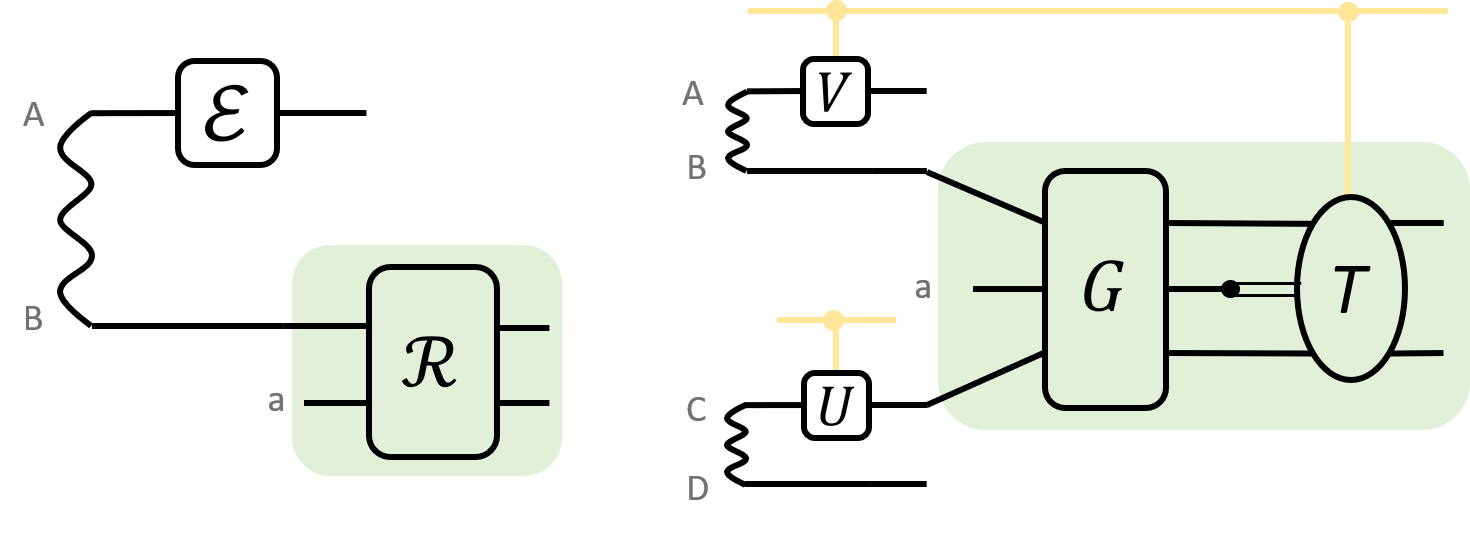}
  \caption{(Left):
  Schematics of the quantum circuit for channel execution of a channel $\C E$
  by the execution scheme $\C R$ (shaded).
  The vertical wavy wire is the entangled Bell state $|\omega\ket$.
  The qubit ancilla a for $\C R$ is the lower wire,
  with initial state $|0\ket$ and then measured.
  The output $\C E(\rho)$ is on the top wire (site A).
  (Right):  Gate teleportation by the indirect Bell measurement $\C G$
  and the conditional rotation $T$ (shaded).
  The middle wire of $\C G$ is the qubit ancilla,
  with initial state $|1\ket$ and then measured.
  The symbol $\bullet\!\!\!=\!\!\!=$ is classical one-control.
  The additional control wires (yellow) are present only for the case of
  nonunitary channels.
  The output state is on sites A and D.}\label{fig:tp}
\end{figure}

\section{Channel-execution from Choi states}
\label{sec:cr}

Now we study our model and we start from the simplest component in it,
which is the readout scheme, i.e.,
the channel-execution step which executes the stored channel on states.

From channel-state duality,
a channel $\C E$ is equivalent to the Choi state
\be \C C := (\C E \otimes \I) (\omega)\ee
for $\omega:=|\omega\ket\bra \omega|$ with $|\omega\ket:=\sum_i |ii\ket/\sqrt{d}$
as a maximally entangled state,
also called a (generalized) Bell state.
For convenience, we label the sites of $\C C $ as A and B,
and B is the 2nd site. See Fig.~\ref{fig:tp} (Left).
The channel $\C E$ can be shuffled to the 2nd site so that
$\C C = (\I \otimes \C E^t) (\omega)$.
This becomes a spatial reflection symmetry when $\C E^t=\C E$.
For the unitary case, this means $U$ is a symmetric matrix (not necessarily real).
Note that, as a mathematical fact,
the transpose $t$ is inevitable
and it is not a physical operation that to be implemented.

We are interested in how to use $\C C$ as a program state (or `resource')
to realize or simulate $\C E$ on a system, labelled by S.
Below we will ignore the site labels.
Given $\C C$, the channel can be executed or simulated in two ways.
The first scheme, named as `Bell scheme', employs Bell measurement and
the channel is realized as
\be \C E(\rho)= \bra \omega |(\C C \otimes \rho )|\omega\ket,\ee
for the projection $|\omega\ket$ on the system S and the site B as a part of the Bell measurement.
Other projections in the Bell measurement will lead to Pauli byproduct operators $P_i$,
which, however, cannot be pulled out after the channel $\C E$ since they do not commute.
In this scheme, we see that the state $\rho$ needs to be prepared initially.

By treating state $\rho$ as a measurement element,
the channel can also be realized as
\be \C E(\rho)= \T tr_B [\C C (\I \otimes \rho^t)], \ee
which can be viewed as a measurement of $\sqrt{\rho^t}$ on site B.
Now the output is on site A.
We name this scheme as `POVM scheme' since here the state $\rho$ serves as a POVM element.
Also note that in practice an algorithm may require output as $\C E(\rho)$ or measurement results on it.
The measurement step on $\C E(\rho)$ is standard as in the circuit model.

We find this scheme can execute the channel in a heralded way instead of being probabilistic,
see Fig.~\ref{fig:tp} (Left).
Define a channel $\C R$ by two Kraus operators
\be K_0= \sqrt{\rho^t},\; K_1=\sqrt{\I-\rho^t}, \label{eq:R}\ee
and the two POVM effects are nothing but $\rho^t$ and $\I-\rho^t$.
The channel $\C R$ can be easily realized by a unitary $U_\C R$ from dilation method,
which takes the form
\be U_\C R=\begin{pmatrix} \sqrt{\rho^t} & \sqrt{\I-\rho^t} \\ \sqrt{\I-\rho^t} & -\sqrt{\rho^t}\end{pmatrix}. \ee
As $[K_0,K_1]=0$, with the eigen-decomposition $\rho^t=VDV^\dagger$,
so $K_0=V \sqrt{D} V^\dagger$, $K_1=V \sqrt{\I-D} V^\dagger$.
The channel is realized as
\be \C R (\sigma)= \T tr_a \C V \circ \C M \circ \C V^\dagger (\sigma\otimes |0\ket\bra 0|),\ee
with the final trace on the qubit ancilla a, which is initialized at $|0\ket$.
The unitary $\C M$ is a `multiplexer' that realizes $\sqrt{D}$ and $\sqrt{\I-D}$,
obtained from the well-known sine-cosine decomposition~\cite{WS15}.
Note here $\sigma$ is not $\rho$, and $\C V(\cdot):=V(\cdot)V^\dagger$ acts on the system.

The ancilla a needs to be measured selectively, i.e., the outcome is recorded.
When it is 0 (1), $K_0$ ($K_1$) is realized on site B,
leading to $\C E(\rho)$ ($\C E(\I-\rho)$) on site A.
In order to make measurement of a certain observable on the final state $\C E(\rho)$,
we have to correct the case of being 1.
The idea is to use another copy of $\C C$ and realize $\C E(\I)$ on site A by tracing out site B.
Then we subtract measurement results on $\C E(\I)$ and $\C E(\I-\rho)$,
which is just the measurement results on $\C E(\rho)$.
When the channel is unitary $U$,
the correction step can be omitted since the effect $\C U(\I)$ is trivial.
To summarize, in the POVM scheme, two copies of $\C C$ are needed in general.
When the outcome on the first copy is 0, then the second copy is not needed;
when it is 1, then the second copy is needed to make the correction.
As such, this scheme is heralded.
Compared with the Bell scheme, it is deterministic by using two samples of the Choi states,
and we will employ it in the stored-program model.

\section{Composition of Choi states and generalized gate-teleportation}
\label{sec:comp}

In our model, a Choi state serves as a program state.
Given two program states, one primary task is to see if there is a composition of them.
The composition is important so that programs (`gates') can be shuffled around and combined together.
To achieve this, we develop a generalized gate teleportation scheme.

\subsection{Unitary case}

We first study the unitary case, for which the Choi state is pure.
Denote two Choi states as $|\psi_U\ket$ and $|\psi_V\ket$,
and the sites as A, B, and C, D, respectively. See Fig.~\ref{fig:tp} (Right).
To simulate gates $UV$ or $VU$, we need to obtain the state $|\psi_{UV}\ket$ or $|\psi_{VU}\ket$.
A standard way is to use the Bell scheme and when the projection is $\omega$ on sites B and C,
the gate $V$ is teleported from site A to site C, and the state of sites A and D is $|\psi_{VU}\ket$.
To obtain $|\psi_{UV}\ket$, we shall apply the Bell scheme on sites A and D.
However, when the Bell measurement yields other outcomes,
there will be Pauli byproduct operators in between the two unitary gates, such as $UP_iV$,
which cannot be corrected in general.

Our method to resolve the difficulty is to replace the Bell scheme by an indirect Bell measurement
which requires an extra qubit ancilla.
First, we call the Bell state $|\omega\ket$ as `singlet', and other threes
as `triplet' for the qubit case~\footnote{
This differs from standard notions which take the antisymmetric one as singlet,
and the three symmetric ones as triplet.
We find this mismatch does not affect our scheme.
The term `triplet' only works for the qubit case.
For qudit case, we use the term `adjointor' since it is the adjoint representation.
The triplet is the adjoint representation of $SU(2)$.
However, we will use the most familiar term `triplet' most of the time.}.
Recall that in the Bell measurement
the circuit to prepare Bell states is firstly run backwards,
denoted as $U_B^\dagger$, and then
projective measurement is performed in the computational basis~\cite{NC00}.
Our method is to use a gate $U_\text{T}$ after $U_B^\dagger$ to couple to the ancilla at state $|1\ket$
and measure the ancilla instead.
Such an indirect or generalized Bell measurement can be expressed as
\be \C G (\sigma) := \T tr_a \; \C U_\text{T} \circ \C U_B^\dagger (\sigma\otimes |1\ket\bra 1|).\ee
The gate $U_\text{T}$ is the Toffoli gate but with one-controls replaced by zero-controls,
and with the qubit ancilla as target.
This encodes the information of being singlet or triplet to the ancilla:
when the outcome on the ancilla is 0 (1), it means the singlet (triplet) is realized.
Note that the singlet case flips the ancilla state.
The measurement on the ancilla can also be pushed to the end:
the classical control will be replaced by a quantum control.
Here we choose to use the classical control.

When the outcome is the singlet on sites B and C,
the gate $U$ on site C is teleported to site A, leading to $|\psi_{VU}\ket$.
When the outcome is the triplet (or adjointor in general),
different from the Bell scheme,
we need to use the symmetry property of the triplet:
it has the full symmetry of $SU(2)$ (or $SU(d)$ in general).
This is because
the channel $\C G$ conditioned on outcome 1 is the set of Pauli operators $P_i$,
i.e., a Pauli channel.
We know from Section~\ref{sec:pre} above that it has the symmetry $SU(d)$ in general.

Now if we intend to shuffle the gate $V$,
we can apply the corresponding rotation $T$ on the triplet,
and this is equivalent to the action of $V^t$ on site C.
This finally leads to the state $|\psi_{V^tU}\ket$.
That is to say, the generalized gate-teleportation scheme can
realize the state $|\psi_{VU}\ket$ or $|\psi_{V^tU}\ket$ in a heralded way
depending on the outcome from the qubit ancilla.
Furthermore,
combined with the POVM scheme on site D for execution,
the gate $VU$ or $V^tU$ can be simulated with output on site A for final readout.

Now we encounter the issue of transpose.
It seems the transpose is inevitable,
which forbids a deterministic gate teleportation (e.g., for $V$).
It turns out this is only apparent and can be avoided by noticing that
any unitary operator can be expressed as a product of two symmetric unitary operators.
Namely, given a unitary operator $U$,
from eigen-decomposition we have $U=U_D D U_D^\dagger$ for unitary $U_D$ and diagonal $D$ matrices.
It can also be written as
\be U= (U_D U_D^t) (U_D^* D U_D^\dagger), \ee
which is a product of two symmetric unitary matrices, denoted as $U^L$ and $U^R$, and $U=U^LU^R$.
This means that for any unitary $U$ of arbitrary dimension,
it only requires two Choi program states, $|\psi_{U^L}\ket$ and $|\psi_{U^R}\ket$,
to deterministically teleport $U$.

In addition, we also believe that using only one Choi state cannot achieve deterministic teleportation.
One might intend to replace the Pauli channel and symmetry group $SU(d)$
by other channels or groups.
This shall need a channel that has a symmetry of the form $U(\cdot) U^*$.
This channel is equivalent to a bipartite symmetric state invariant under $U\otimes U^\dagger$~\cite{Wer89,HH99},
which turns out to be the trivial identity state.
This is nothing but the case when a singlet is established in teleportation.
This shows that the composition method above has to be used.
A different approach is to allow non-Pauli byproduct, like the Clifford hierarchy~\cite{GC99},
which is beyond our framework based on symmetry,
and this also requires many Bell states (hence program states) especially for gates on higher levels of the hierarchy.

The gate teleportation above directly extends to the composition of a sequence of program states.
The whole sequence of gates obtained are of the form $P \prod_\ell U_\ell$,
for $P$ as the final Pauli byproduct,
and each gate is $U_\ell=U_\ell^L U_\ell^R$,
and corresponding rotations $T_\ell$ are needed for the adjointor case.

\subsection{General cases}

We now study the case for channels whose Choi states are mixed states.
The symmetry condition does not generalize directly since now unitary operators
are replaced by nonunitary quantum channels.
This appears as a nontrivial obstacle,
while we find that a scheme based on direct-sum dilation shall work.
For a channel $\C E$ with a set of Kraus operators $\{K_i\}$,
each of them can be dilated to a unitary
\be U_{K_i}=\begin{pmatrix} K_i & \sqrt{\I-K_i^\dagger K_i} \\ \sqrt{\I-K_i K_i^\dagger} & -K_i^\dagger\end{pmatrix}. \ee
The gate $U_{K_i}$ acts on a space of dimension $2d$.
Denote the original space as $\C H_S$, and the additional one as $\C H_S^\perp$.
Now the channel $\C E$ can be simulated by a random-unitary channel
\be \C E'(\sigma)= \T tr_a \; \breve{\C U} (\sigma \otimes e) \ee
for $\breve{U}:=\sum_i |i\ket\bra i| \otimes U_{K_i}$ as a controlled-unitary gate,
with the ancilla as control with initial state $e:=|e\ket\bra e|$ and traced out,
for $|e\ket:=\sum_i |i\ket$ as the equal-amplitude superposition state.
The state $\sigma=\rho\oplus \mathbf{0}$, for $\mathbf{0}$ on the dilated subspace $\C H_S^\perp$.
The action $\C E(\rho)$ is the restriction of $\C E'(\sigma)$ to the system subspace $\C H_S$.

Now compare with the unitary case,
here the task is to teleport controlled-unitary gates instead of unitary gates.
This can be done using a slight extension of our scheme above:
for each $U_{K_i}$ in $\breve{U}$, there exists a $T_i$ gate that can teleport it.
The $T_i$ gates are controlled by the same ancilla for $U_{K_i}$.
See Fig.~\ref{fig:tp} (Right), the yellow control wires.
Similar with the unitary case,
when we obtain the singlet, the channel $\C E$ is teleported.
To avoid the transpose,
we decompose each $U_{K_i}$ as the product of two symmetric unitary matrices
$U_{K_i}=U^L_{K_i}U_{K_i}^R$,
and then it is not hard to see that, using the same control wire for $U^L_{K_i}$ and $U_{K_i}^R$
and using two program states,
the gate $\breve{U}$ can be teleported,
i.e., the channel $\C E$ is teleported.

To execute the action of channel on state,
we need to design a POVM and a channel based on state of the form $\rho\oplus \mathbf{0}$.
The channel, denoted $\C R'$, contains three Kraus operators
\be K_0= [\sqrt{\rho^t},\mathbf{0}], \; K_1=[\sqrt{\I-\rho^t},\mathbf{0}], \;
K_2= [\mathbf{0},\I],\label{eq:Rp}\ee
as an extension of the channel $\C R$~(\ref{eq:R}).
This channel needs a qutrit ancilla, and differently from $\C R$,
when the outcome is 2,
the simulated result restricted to the space $\C H_S^\perp$ is $\C E^\dagger(\I)$,
which equals $\I$ from the trace-preserving condition.
If this occurs, the simulation has to be started again.
We observe that the direct-sum dilation suits some models and systems.
For instance, in linear optics and quantum walk,
each computational basis state can be addressed by encoding in a separate `mode',
and a direct-sum dilation is to add more modes to the system.
As such, the restriction to a subspace can be easily done by only observing those modes in it
and the scheme is deterministic.

For special type of channels, our scheme can be greatly simplified.
A wide class of channel are the random unitary channels,
and it is clear that they can be realized by the controlled-unitary scheme above,
and no direct sum dilation is needed.
For qubit channels, all unital channels, which preserve the identity,
are random unitary channels, hence can be easily simulated.
Another kind is known as entanglement-breaking channels~\cite{Rus03},
for which the Choi states are bipartite separable states
(while the partial traces are $\C E(\I)$ and $\C E^t(\I)$).
These channels and program states would be trivial since there is no entanglement and
they can be easily simulated by a measurement-preparation scheme.

\section{Preparation of program states via generalized-extreme channels}
\label{sec:extr}

Now we study the preparation of program states if they are not given for free.
A Choi state $\C C$ is not easy to prepare on the first hand, namely,
this may require the operation of $\C E$ on the Bell state $|\omega\ket$,
and realizing $\C E$ itself (e.g., by a dilated unitary) is a nontrivial task.
From Stinespring's dilation, we know that it requires the form of Kraus operators,
which are not easy to find in general given a Choi state.
Here we develop a framework that relies on Choi states and
reduces the quantum circuit complexity by using random bits.

The set of qudit channels forms a convex body.
This means that a convex sum of channels still leads to a channel,
and there are extreme channels that are not convex sum of others.
From Choi~\cite{Cho75}, a channel is extreme iff there exists a Kraus representation $\{K_i\}$
such that the set $\{K_i^\dagger K_j\}$ is linearly independent.
For a qudit, this means the rank of an extreme channel is at most $d$.
Channels of rank $r\leq d$ are termed as generalized-extreme channels~\cite{RSW02},
here termed as `gen-extreme channels.'
It is clear to see that, a gen-extreme but not extreme channel
is a convex sum of extreme channels of lower ranks.
It has been conjectured~\cite{Rus07} and numerically supported~\cite{WS15,Wang16} that
arbitrary channel can be written as a convex sum of at most $d$ gen-extreme channels
$\C E=\sum_{i=1}^d p_i \C E_i^g$.
This requires a random dit.
For the worst case,
the upper bound for such a convex sum is $d^4-d^2$ from Carath\'{e}odory theorem on convex sets~\cite{GW93},
which merely costs more random dits.

To simulate the composition $\prod_i \C E_i$,
with each $\C E_i$ of rank greater than $d$, hence permitting a convex-sum decomposition,
one needs to sample the composition of gen-extreme channels.
We find there exists a concise form of Choi states for gen-extreme channels,
which can be used to find the circuit and also Kraus operators directly.

The Choi state $\C C$ for a gen-extreme channel $\C E$ is of rank $r\leq d$ and $\T tr_A\C C=\I$,
$\T tr_B\C C=\C E(\I)$.
It turns out $\C C=\sum_{ij} |i\ket \bra j| \otimes \C C_{ij}$ for
\be \C C_{ij}:=\C E^t(|i\ket \bra j|) = \sqrt{\C C_{i}}U_i^\dagger U_j \sqrt{\C C_{j}}\ee
for $\C C_{i}\equiv \C C_{ii}$, and $U_i, U_j\in SU(d)$~\cite{WS15}.
Observe that \be \C E^t(\rho)=\sum_{ij}\rho_{ij} \C C_{ij}=V^\dagger (\rho \otimes \I) V, \ee
for an isometry $V:=\sum_i |i\ket U_i \sqrt{\C C_{i}}$.
Here $\I$ is an ancilla state.
Now we show $V$ can be used to find a quantum circuit to realize $\C E$.
Given $V$, we can find a unitary dilation $U$ such that $U|0\ket=V$,
and it relates to the channel by $\C E^t(\rho)=\bra 0| U^\dagger (\rho \otimes \I) U |0\ket $,
while the final projection $|0\ket\bra 0|$ is on the system.
Define $W:=\text{swap}\cdot U^\dagger$ for swap gate between the system and ancilla which are of the same dimension,
then we find
\be \C E(\rho)=\T tr_a \C W^t (\rho\otimes |0\ket\bra 0|), \ee
which means $W^t$ is the circuit to realize the channel $\C E$ as in the standard dilation scheme.
The Kraus operators can be obtained from it as $K_i=\bra i|W^t|0\ket$.

Compared with standard (tensor-product) dilation method to simulate a general channel~\cite{NC00},
which will require two qudit ancilla,
the method above requires lower circuit cost since it only needs a single qudit ancilla instead of two.
While the convex-sum decomposition, which is a sort of generalized eigenvalue decomposition
since a gen-extreme Choi state can be mapped to a pure state,
are difficult to solve for large-dimensional channels,
it shall be comparable with the eigen-decomposition of Choi state
to find the set of Kraus operators.
Both of the decompositions are solvable for smaller systems.

\section{Discussions}
\label{sec:app}

In the above we have discussed the primary components in our model.
We did not study fault tolerance which is not the goal of this model.
Instead, we assume the qubits, gates, and measurements are fault-tolerant,
which can be achieved with quantum error-correcting codes.
Here we discuss two interesting issues
to make connections with some standard framework or results.

\subsection{Teleportation of universal gate set}

A computation is universal if the group $SU(2^n)$ can be realized for any integer $n$.
The common approach to achieve universality is by gate-compiling
based on universal gate sets~\cite{DN06}.
Our method can be used to teleport unitary universal gate sets.
Consider the popular Hadamard gate H, phase gate S, the so-called $Z^{1/4}$ gate T, CNOT, CZ, and Toffoli gate.
One immediately notice that
these gates are all symmetric matrices.
We see above that symmetric unitary operators, for which $U=U^t$,
can be teleported deterministically,
and the byproduct are Pauli operators.
Note a product of symmetric matrices are not symmetric in general.

It is easy to check that the affine forms of H, S, CNOT, and CZ
are (generalized) permutations since they are Clifford gates which preserve the Pauli group.
A generalized permutation is a permutation which also allows entry of module 1 besides 1 itself.
The T gate and Toffoli gate are not Clifford gates, and their affine forms are not permutations.
Instead, the affine forms of them contains a Hadamard-like gate as a sub-matrix,
which means, in the Heisenberg picture, they are able to generate superpositions of Pauli operators.
This fact also generalize to the qudit case,
with Hadamard replaced by Fourier transform operators.
This serves as an intriguing fact regarding the origin of
computational power of quantum computing.

\subsection{Stored-program quantum computing}

In their original model,
it is proved that the operation $U|d\ket$ for general $U$ and input data state $|d\ket$
cannot be simulated by $G|d\ket|\psi_U\ket$ in a unitary way,
for $G$ independent of $U$ and $|d\ket$~\cite{NC97}.
Our method serves as a modification of it so this is possible:
the processor $G$ is the generalized gate teleportation scheme,
which instead depends on the input program state.
For symmetric matrices $U$, the program state $|\psi_U\ket$ is sufficient.
For general cases, with $U=U^LU^R$, the program states
$|\psi_{U^L}\ket$ and $|\psi_{U^R}\ket$ together are sufficient to obtain
deterministic teleportation and composition.
The channel-execution scheme does not depend on the input program,
which, however, is destroyed after the computation.
One has to either prepare multiple copies of the program states
or refresh the program if required.
This is similar with the `one-way' character of measurement-based quantum computing~\cite{RB01},
and indeed, our model can be treated as a generalization of it as we have mentioned.
By the composition of a sequence of program states,
we prepare many-body states in a heralded way,
which can be properly treated as modification of graph states~\cite{BBD+09}
or valence-bond solid states~\cite{AKLT87}.

\section{Conclusion}
\label{sec:conc}

In this work, we have studied computational schemes based on Choi states,
which serve as the program states in a stored-program model.
Meanwhile, Our model can be viewed as an enrichment of measurement-based computing
by using general Choi states and their compositions.
With a symmetry point of view of teleportation and computation,
we develop a symmetry-based generalization of gate-teleportation,
which is shown to be powerful to deterministically build up networks of program states.
Our study reveals that quantum stored-program computing or information processing is powerful
and shall be pursued further.

A program, as an operator being either unitary or nonunitary,
can be stored by one Choi state if it is a symmetric matrix or two Choi states in general.
For nonunitary channels, i.e., mixed Choi program states,
we first embed each nonunitary Kraus operators into a unitary one based on direct-sum dilation method,
which is further treated as a symmetry operator.
Whether channels can be directly treated as a `quasi-symmetry' or not is an interesting open problem.
In addition, we discussed two schemes for the execution of a channel from its Choi state:
one is probabilistic, and the other is heralded.
In practice, this may cause various complications for different tasks.
This reveals some intriguing characters or limitations of Choi state formalism,
and it is not clear if any improvement is possible.

The framework we develop may also be generalized.
We expect that it can be adopted to infinite-dimensional systems
for the teleportation of continuous variables.
Our scheme can also be used in many settings such as
universal computing models, quantum simulation,
and quantum communication.

\section{acknowledgement}
This work was funded by NSERC and ISED of Canada.
Discussions with O. Kabernik, R. Raussendorf, and Y.-J. Wang are acknowledged.

\newpage

\bibliography{ext}

\end{document}